\documentclass[11pt]{article}
\usepackage[hyperref]{acl}
\usepackage{times}
\usepackage{latexsym}
\usepackage{booktabs}
\usepackage{multirow}
\usepackage{graphicx}
\usepackage{amsmath}
\usepackage{url}
\usepackage{tikz}
\usepackage{microtype}
\usepackage{enumitem}

\let\oldbibliography\thebibliography
\renewcommand{\thebibliography}[1]{%
  \oldbibliography{#1}%
  \setlength{\itemsep}{1.5pt}%
  \setlength{\baselineskip}{11.25pt}
  \setlength{\lineskiplimit}{-\maxdimen}
}

\usetikzlibrary{positioning}

\title{Perceptually Better, Semantically Worse:\\
Measuring Speech Enhancement Impact on
LLM-Based Voice Systems}

\author{
  Randy Frans Fela {\normalfont and}
  Pejman Mowlaee \\
  GN Group, Ballerup, Denmark\\
  \texttt{\{rffela, pmowlaee\}@gn.com}
}

\begin{document}
\maketitle

\begin{abstract}
Speech enhancement (SE) is commonly applied as a preprocessing step
in spoken AI pipelines under the assumption that better audio quality
improves downstream task performance. Whether SE-induced distortions
propagate to downstream LLM task performance remains an open question.
We introduce Output Divergence Rate (ODR), which measures how often SE
changes an LLM's intent classification relative to clean speech, and
benchmark five conditions on 2{,}974 SLURP clips using Whisper
large-v3 and wav2vec2-large cascades. Every condition produces ODR significantly above zero
($p < 0.001$, binomial test). MetricGAN{+}
more than doubles ODR versus unenhanced noisy speech (0.318 vs.\
0.135) despite improving PESQ, and unmitigated echo reaches an ODR of
0.836 through speaker substitution, a failure WER
cannot capture. Audio quality metrics range from near-zero to moderate correlation with ODR
(SQUIM-MOS $\rho\!=\!-0.068$, PESQ $\rho\!=\!-0.467$).
The MetricGAN{+} and echo results replicate across ASR
architectures, indicating that standard audio quality metrics are
insufficient for LLM pipeline quality.
\end{abstract}
 
\section{Introduction}
\label{sec:intro}
 
Modern spoken AI systems chain three components: speech enhancement
(SE), automatic speech recognition (ASR), and
large language model (LLM).
The SE step is assumed acoustically neutral, yet recent evidence
challenges this at the transcription level. \citet{chondhekar2025noising}
show MetricGAN{+} increases semWER across all 40 tested ASR
configurations, and \citet{islam2026denoising} report the same for
diffusion-based SE.
What remains unknown is whether these distortions propagate to LLM
reasoning. This question is not straightforward: LLMs exhibit
robustness to noisy text~\citep{chen2024voicebench, chen2026voicebench},
so transcription degradation may not translate to semantic task failure,
or it may cause different failures that WER cannot measure.
 
In this paper, we propose \textbf{Output Divergence Rate (ODR)},
a clean-reference task-output divergence metric for measuring
SE-induced instability in cascaded ASR--LLM voice systems,
benchmarked across five enhancement
conditions on SLURP~\citep{bastianelli2020slurp}.
Results show that LLMs absorb transcription errors to varying
degrees (WER--ODR gap $< 0$ for all conditions), yet echo
conditions produce catastrophic ODR through a qualitatively
distinct speaker-substitution failure that WER cannot capture.
Standard audio quality metrics do not predict this damage,
and findings are consistent across two architecturally distinct ASR
models, ruling out Whisper-specific artefacts. Our main contributions are as follows:

\begin{table*}[t]
  \centering
  \begin{tabular}{lll}
    \hline
    \textbf{Condition} & \textbf{Category} & \textbf{Description} \\
    \hline
    Clean      & Reference       & Original SLURP recording \\
    Noisy      & Degraded        & DNS noise, SNR\,=\,10\,dB \\
    MetricGAN+ & Discrim.\ SE    & GAN noise suppression \\
    Echo (sim) & Echo            & Simulated echo, no AEC \\
    Echo + DTLN-AEC & AEC        & DTLN-AEC on echo sim. \\
    Dereverb   & Dereverberation & WPE (5 iterations) \\
    \hline
  \end{tabular}
  \caption{Enhancement conditions applied to 2{,}974 SLURP test
    recordings. All degradations simulated using DNS Challenge corpora.}
  \label{tab:conditions}
\end{table*}

\begin{enumerate}[itemsep=1.5pt, topsep=2pt, parsep=0pt, partopsep=0pt]
\item We propose \emph{Output Divergence Rate} (ODR), a
  task-level metric that measures how often speech enhancement
  changes the downstream LLM prediction relative to clean speech.
\item We construct a six-condition benchmark spanning noise
  suppression, echo cancellation, and dereverberation on
  2{,}974 SLURP test clips with 77 intent classes, and release
  the full pipeline at \href{https://github.com/fransfela/se-llm-odr-benchmark}{https://github.com/fransfela/se-llm-odr-benchmark}.
\item We show that enhancement can introduce systematic intent
  collapse toward catch-all categories, a failure mode
  invisible to WER and standard quality metrics.
\item We show that standard audio quality metrics reflect
  between-condition ODR separation but have near-zero predictive
  power for which individual clips will diverge once the enhancement
  condition is fixed, making them unreliable per-clip early-warning
  signals.
\item Findings replicate across two ASR architectures
  (attention-based and CTC-based) and are robust to LLM capacity, 
  indicating condition ranking and primary conclusions are
  architecture- and classifier-agnostic.
\end{enumerate}

\section{Related Work}
\label{sec:related}

SE-induced ASR degradation is well documented.
\citet{lee2023d4am} propose joint SE+ASR training to mitigate it.
\citet{chondhekar2025noising} show that MetricGAN{+} increases semWER
across all tested configurations, and \citet{islam2026denoising}
report the same pattern for diffusion-based SE.
None of this work asks whether the resulting transcription errors
actually change what a downstream LLM does with the output, which
is the question we address here.

At the downstream NLP level, \citet{rai2024denoasr} show that
denoising amplifies demographic bias in ASR, and \citet{ali2022time}
propose joint SE+intent training, noting that independently optimised
SE may not serve downstream NLP well.
Their solution requires retraining the SE model jointly with the
downstream task, while we focus on diagnosing this effect in
pipelines as they are already deployed, without retraining anything.

Closest to our work, VoiceBench~\citep{chen2024voicebench, chen2026voicebench}
evaluates LLM voice assistants under acoustic variation but treats the
whole audio-to-response pipeline as a black box rather than isolating
the SE stage.
The URGENT challenge~\citep{zhang2024urgent} benchmarks SE across a
wide range of distortions using perceptual and ASR-based metrics.
Its 2025 edition~\citep{saijo2025interspeech} introduced
downstream-task metrics (word accuracy, speaker similarity) but
no LLM semantic output evaluation.
Our benchmark fills this gap by isolating SE as the variable of
interest and measuring its effect on an LLM's task output rather
than on transcription or signal quality alone.
 
\section{Benchmark Design}
\label{sec:method}
 
\subsection{Dataset and Conditions}
 
We use the \texttt{slurp\_real} test split of
SLURP~\citep{bastianelli2020slurp}: 2{,}974 real human recordings
across 18 domains and 77 intent classes.
The TTS subset is excluded as its uniform acoustics suppress
enhancement artefacts.
Table~\ref{tab:conditions} summarises the five enhancement conditions,
chosen to cover the major SE classes deployed in commercial spoken AI.

\textbf{Noisy} adds DNS Challenge noise~\citep{reddy2021interspeech} at
SNR$\,=\,$10\,dB without enhancement, establishing a degraded
baseline.
\textbf{MetricGAN{+}}~\citep{fu2021metricgan+} applies discriminative
GAN-based noise suppression trained to maximise PESQ, representative
of the most widely deployed SE class.
\textbf{Echo~(sim)} mixes near-end speech with a DNS far-end signal
convolved through a DNS room impulse response at a
signal-to-echo ratio $\text{SER}\!\in\!\{-10,\,-5,\,0\}$\,dB, without
any cancellation, isolating a conferencing-specific failure mode
absent from prior NLP benchmarks.
\textbf{Dereverberation} applies WPE~\citep{nakatani2010speech} to
reverberant speech ($\text{RT60}\!\in\!\{0.3,\,0.6,\,0.9\}$\,s).

\textbf{Echo + DTLN-AEC} applies DTLN-AEC~\citep{westhausen2021acoustic}
(512-unit, 10.4M parameters), a dual-signal transformation LSTM
network for acoustic echo cancellation, to the Echo~(sim) microphone
signal using the far-end reference signal.
This condition tests whether neural AEC restores LLM
performance degraded by unmitigated echo.
Figure~\ref{fig:spectrograms} (Appendix~\ref{app:spectrograms}) shows
log-mel spectrograms of representative utterances, illustrating the acoustic signature of all six conditions.
 
\subsection{Evaluation Pipeline}
 
Each clip is transcribed by two ASR models, Whisper
large-v3~\citep{radford2023robust} (encoder-decoder, beam size 5,
temperature 0) as the primary model and wav2vec2-large-960h~\citep{baevski2020wav2vec}
(self-supervised CTC) as the architectural comparison.
Whisper WER on clean SLURP test speech is 0.507, reflecting the
challenging spontaneous-speech acoustics of the corpus~\citep{bastianelli2020slurp}.
All ODR values are measured relative to this same clean baseline.

\begin{table*}[t]
  \centering
  \resizebox{2.05\columnwidth}{!}{%
  \begin{tabular}{lrrrrrrrrrr}
    \hline
    \textbf{Condition} & \textbf{PESQ} & \textbf{STOI} &
    \textbf{SNR} & \textbf{SI-SDR} &
    \textbf{SRMR} & \textbf{SQUIM} & \textbf{WER} & \textbf{WER$_\text{cap}$} & \textbf{ODR} \\
    \hline
    Clean           & 4.64 & 1.00 & 105.12 & 94.35 & 8.56 & 4.22 & 0.507 & 0.507 & -- \\
    Noisy           & 1.76 & 0.87 & 9.92 & 10.00 & 6.24 & 3.77 & 0.531 & 0.520 & 0.135 \\
    MetricGAN+      & 1.94 & 0.80 & 1.81 & 4.23 & 10.96 & 3.87 & 0.647 & 0.614 & \textbf{0.318} \\
    Echo (sim)      & 1.11 & 0.52 & $-$2.69 & $-$5.01 & 6.24 & 3.69 & \textbf{1.448} & 0.928 & \textbf{0.836} \\
    Echo + DTLN-AEC & 1.53 & 0.47 & $-$1.66 & $-$26.13 & 10.98 & 3.92 & 0.700 & 0.649 & 0.404 \\
    Dereverb        & 2.43 & 0.71 & $-$4.72 & $-$23.04 & 7.79 & 3.94 & 0.517 & 0.507 & 0.108 \\
    \hline
  \end{tabular}%
  }
  \caption{Mean audio quality metric scores, Whisper WER, and ODR per
    condition (Clean shown as reference, ODR\,=\,0 by definition).
    WER: raw mean per-clip word error rate.
    WER$_\text{cap}$: mean of $\min(\text{WER}_i, 1.0)$ per clip,
    used in the WER--ODR gap (Eq.~\ref{eq:gap}).
    Bold marks the highest ODR values
    (Echo~(sim) and MetricGAN{+}) and the saturating raw WER
    (Echo~(sim) 1.448).}
  \label{tab:metrics}
\end{table*}

Transcripts are classified by Gemini~2.5 Flash Lite via the Gemini
API using a fixed closed-set prompt over the 77 SLURP intents
(temperature 0, max 15 tokens, invalid rate 0.07--0.20\% across
conditions with no systematic condition dependence, see Appendix~\ref{app:prompt}).
ODR for condition $C$ is:
\begin{equation}
\text{ODR}(C) = \frac{|\{i : \hat{y}_C^{(i)} \neq
\hat{y}_{\text{clean}}^{(i)}\}|}{N}
\label{eq:odr}
\end{equation}
where $\hat{y}_C^{(i)}$ is the predicted intent under condition $C$,
$\hat{y}_{\text{clean}}^{(i)}$ is the clean baseline, and $N$
excludes invalid predictions.
We additionally report the WER--ODR gap:
\begin{equation}
\text{Gap}(C) = \text{ODR}(C) - \frac{1}{N}\sum_{i=1}^{N}\min(\text{WER}_C^{(i)},\,1.0)
\label{eq:gap}
\end{equation}
where $N$ is the same denominator as in Eq.~(\ref{eq:odr}).
WER is clipped per utterance before averaging to bound the
contribution of speaker-substitution clips whose raw WER
exceeds 1.0 (Echo (sim): mean raw WER\,=\,1.448).
A negative gap indicates the LLM is partially robust to the
transcription errors induced by that condition.
We report 95\,\% bootstrap CIs ($B\!=\!10{,}000$, seed$\,=\,$42).
 
\subsection{Quality Metrics}
 
We compute six metrics spanning intrusive perceptual (PESQ, STOI),
intrusive signal-level (SNR, SI-SDR), and non-intrusive MOS
prediction families (SQUIM-MOS, SRMR), and report Spearman $\rho$
and Pearson $r$ against binary ODR with Bonferroni correction
($n\!=\!6$).
 
\section{Results}
\label{sec:results}
 
\subsection{ODR Across Conditions}
 
\begin{figure}[t]
\centering
\includegraphics[width=\columnwidth]{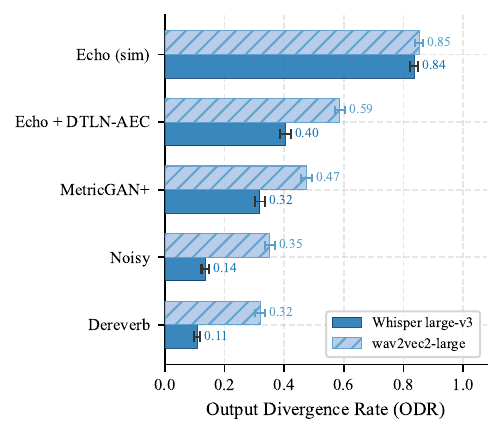}
\vspace{-.75cm}
\caption{ODR by condition for Whisper large-v3 and wav2vec2-large,
with 95\,\% CIs. All conditions are significant ($p\!<\!0.001$,
binomial test). Rankings are identical across architectures.}
\label{fig:odr}
\end{figure}

Figure~\ref{fig:odr} reports ODR for both
ASR models across all non-clean conditions.
All conditions produce statistically significant ODR ($p\!<\!0.001$),
and the condition ranking is identical for both models
(Echo (sim) $>$ Echo + DTLN-AEC $>$ MetricGAN{+} $>$ Noisy $>$ Dereverb).

Under Whisper, MetricGAN{+} produces ODR of 0.318
(CI\,[0.301,\,0.335]), more than doubling the unenhanced noisy
baseline (0.135, CI\,[0.123,\,0.147]).
MetricGAN{+} raises PESQ yet simultaneously worsens STOI and more
than doubles ODR, demonstrating that perceptual quality improvement
and LLM task performance are not aligned.
Dereverberation produces the lowest ODR (0.108,
CI\,[0.097,\,0.119]), comparable to the noisy baseline.
Under wav2vec2 the same ordering holds at higher absolute values
(MetricGAN{+}: 0.474, noisy: 0.352), consistent with CTC decoding's
greater sensitivity to frame-level spectral artefacts.

\begin{table*}[t]
  \centering
  \begin{tabular}{llll}
    \hline
    \textbf{SLURP ID} & \textbf{Condition} & \textbf{Transcript} &
    \textbf{Intent} \\
    \hline
    11862 & Clean & ``Recommend a movie for me.'' & \texttt{recommendation\_movies} \\
     & MetricGAN{+} & ``Thank you.'' & \texttt{general\_greet} \\
    \hline
    14746 & Clean & ``What is the capital of Kazakhstan?'' & \texttt{qa\_factoid} \\
     & MetricGAN{+} & ``This is a caption for Justin.'' & \texttt{general\_quirky} \\
    \hline
    12308 & Clean & ``Book me a train ticket.'' & \texttt{transport\_ticket} \\
     & MetricGAN{+} & ``We're free to go.'' & \texttt{general\_quirky} \\
    \hline
  \end{tabular}
  \caption{Example clips where MetricGAN{+} causes intent
    collapse to catch-all categories. Transcription remains
    fluent but loses domain-specific lexical cues, causing the
    LLM to default to generic intents. SLURP IDs refer to
    \texttt{slurp\_real} test recordings~\citep{bastianelli2020slurp}.}
  \label{tab:examples}
\end{table*}

Table~\ref{tab:metrics} quantifies the
perceptual--semantic decoupling.
MetricGAN{+} raises PESQ from 1.76 to 1.94
relative to Noisy ($\Delta\!=\!+0.18$), a modest gain that keeps
both values within the ``bad'' range of the 1--4.5 scale,
while simultaneously \emph{reducing}
STOI from 0.87 to 0.80 ($\Delta\!=\!-0.07$) and
more than doubling ODR from 0.135 to 0.318.
This triple divergence is particularly damaging for
practitioners: MetricGAN{+}'s explicit training objective,
PESQ, improves, the intelligibility proxy, STOI, worsens,
and the downstream LLM task failure rate doubles.
No single quality metric provides a reliable signal
about the direction of LLM task impact.

\begin{figure}[t]
\centering
\includegraphics[width=\columnwidth]{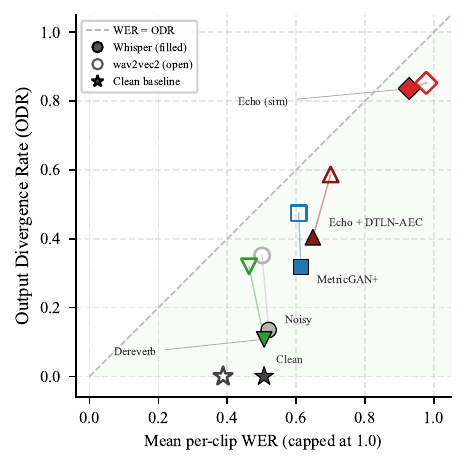}
\vspace{-.75cm}
\caption{WER vs.\ ODR for Whisper and wav2vec2 (filled and open markers).
\textbf{$\star$} marks the clean baseline (ODR\,=\,0 by construction,
WER reflecting Whisper/wav2vec2 accuracy on spontaneous SLURP speech).
All points fall below $y\!=\!x$ for spectral distortion
conditions. This reflects partial LLM robustness to transcription
errors, while Echo~(sim) negative gap reflects WER ceiling
saturation rather than LLM recovery.}
\label{fig:wer_gap}
\end{figure}

\paragraph{A second decoupling: Echo + DTLN-AEC.}
Table~\ref{tab:metrics} shows a different decoupling pattern
for AEC recovery. While DTLN-AEC improves PESQ
(1.11$\rightarrow$1.53), WER ($-51.6\%$), and ODR ($-51.7\%$)
relative to Echo~(sim), SI-SDR worsens sharply
($-5.01\rightarrow-26.13$) and SRMR rises above even the
Clean condition (10.98 vs.\ 8.56).
We attribute this to the gain and phase adjustments DTLN-AEC
applies to suppress the far-end signal, changes immaterial to intent recognition, but register as severe distortion under SI-SDR and as elevated modulation depth under SRMR. STOI also drops below its Echo~(sim) value (0.47 vs.\ 0.52) because STOI's short-time envelope correlation model assumes linear, memoryless distortion, an assumption nonlinear echo cancellation deliberately violates, so STOI scores can fall even when intelligibility improves. SI-SDR, SRMR, and STOI are not designed for AEC evaluation,
where output is intentionally gain- and phase-modified relative to
the near-end reference.
A parallel argument applies to Dereverb: intrusive metrics compare
the WPE output against the dry clean recording. WPE cannot perfectly
invert the room impulse response, so residual differences between
its output and the dry reference produce severely negative SNR
($-$4.72\,dB) and SI-SDR ($-$23.04\,dB) even though ODR is the
lowest of all conditions (0.108), confirming that signal-level metrics
do not track LLM task failure.
Taken together with the MetricGAN{+} case, these results show
that metrics from the same category (intrusive signal-level:
SNR, SI-SDR) can disagree sharply about the direction of an
enhancement's effect, independent of its actual impact on the
downstream task.

\begin{figure*}[ht!]
\centering
\includegraphics[width=0.5\linewidth]{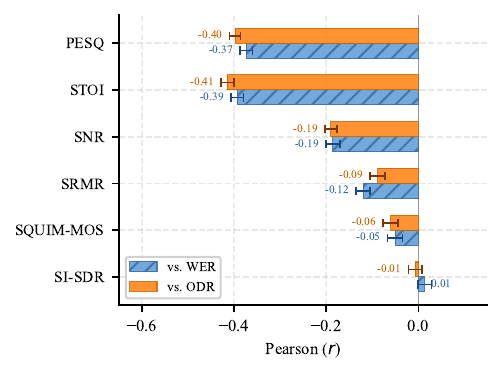}\hfill
\includegraphics[width=0.5\linewidth]{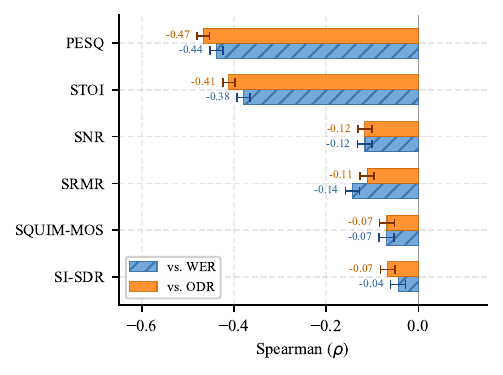}
\vspace{-.75cm}
\caption{Pearson ($r$, left) and Spearman ($\rho$, right) of each
quality metric against WER and ODR. PESQ and STOI
correlate comparably with both targets. Non-intrusive SQUIM-MOS is
near-zero for both, underscoring its inability to predict either
transcription or semantic damage. All Spearman $\rho$ values
significant ($p\!<\!0.001$, Bonferroni corrected). Pearson SI-SDR
is non-significant ($p\!=\!0.40$). Error bars show 95\%
bootstrap confidence intervals.}
\label{fig:correlations}
\end{figure*}

\paragraph{WER--ODR gap.}
Figure~\ref{fig:wer_gap} plots WER against ODR for both models.
All ten points fall below the diagonal ($y\!=\!x$): the LLM is
partially robust to transcription errors in every condition and for
both architectures, with Whisper gaps ranging from $-0.092$ to
$-0.400$ and wav2vec2 gaps from $-0.114$ to $-0.150$.
For Echo (sim), ODR of 0.836 and capped WER of 0.928 yield a
gap of $-0.092$, reflecting WER saturation rather than LLM robustness. With raw
WER of 1.448, Whisper transcribes the far-end loudspeaker rather than
the user, producing a valid but entirely wrong transcript.

\paragraph{Echo condition.}
The echo condition produces ODR of 0.836 (CI\,[0.823,\,0.849]) under
Whisper and 0.853 under wav2vec2, converging across architectures.
At $\text{SER}\!\leq\!0$\,dB, both models transcribe the louder
far-end echo signal rather than the near-end speaker.
This is a categorically different failure from spectral distortion:
the LLM processes a semantically unrelated utterance, and WER cannot
characterise the damage because it saturates.

\paragraph{Echo with cancellation.}
Applying DTLN-AEC neural echo cancellation to the
Echo~(sim) signal substantially recovers ASR
performance, with mean WER dropping from 1.448 to 0.700
($-51.6\,\%$ relative, full WER distribution in
Appendix~\ref{app:wer}), confirming that the echo
canceller correctly suppresses the dominant far-end
signal and restores near-end speech transcription.
The downstream LLM impact follows proportionally:
ODR drops from 0.836 to 0.404
(CI\,[0.386,\,0.422], Gap\,=\,$-0.245$),
a $51.7\,\%$ relative reduction that closely mirrors
the $51.6\,\%$ WER reduction.
This proportionality suggests that the
speaker-substitution artefact is the primary damage
mechanism for echo. Once AEC restores the correct
speaker, both transcription and intent recovery
improve at nearly the same rate.

\paragraph{Systematic intent collapse.}
Of 945 diverged clips under MetricGAN{+}, 48.4\% of the 562
regressions land in \emph{general\_quirky} (13.4\% of all
MetricGAN{+} predictions, vs.\ 5.2\% under clean) or
\emph{general\_greet} (6.8\% vs.\ 2.0\%), both 2.5--3.4$\times$
their clean-speech frequency.
This reveals a distributional shift in LLM input space. SE removes
domain-specific lexical cues, and the LLM defaults to catch-all
categories when the utterance is stripped of its identifying
vocabulary.
Table~\ref{tab:examples} shows representative examples of this
collapse.

\paragraph{Correct vs.\ harmful divergence.}
ODR treats every clean-vs-condition disagreement as damage, conflating
genuine harm with the rarer case in which SE incidentally corrects a
clean-speech misclassification. Using SLURP ground-truth intents, we
decompose divergent clips for each condition into corrections (clean
wrong, condition correct), regressions (clean correct, condition
wrong), and both-wrong shifts (Table~\ref{tab:correction}). Regressions outnumber corrections in
every condition, ranging from 1.9\,:\,1 under Dereverb to
23.1\,:\,1 under Echo~(sim), with Noisy at 3.3\,:\,1, confirming
that ODR predominantly captures harmful divergence rather than
benign correction. MetricGAN{+}'s 7.6\,:\,1
ratio confirms that its intent damage is real, rather than an artefact
of clean-speech classification errors. A McNemar paired test on the
same 2{,}966 clips confirms that MetricGAN{+} causes significantly
more regressions than Noisy ($\chi^2\!=\!427$, $p\!<\!10^{-90}$).

\begin{table*}[t]
  \centering
  \begin{tabular}{lrrrrr}
    \hline
    \textbf{Condition} & \textbf{Correction} & \textbf{Regression} &
    \textbf{Both-Wrong} & \textbf{Unaffected} & \textbf{Regr.\,:\,Corr.} \\
    \hline
    Noisy        & 1.8\,\% & 6.0\,\%  & 5.8\,\%  & 86.5\,\% & 3.3\,:\,1  \\
    MetricGAN+   & 2.5\,\% & 18.9\,\% & 10.4\,\% & 68.2\,\% & \textbf{7.6\,:\,1}  \\
    Echo (sim)   & 2.5\,\% & 58.4\,\% & 22.7\,\% & 16.4\,\% & \textbf{23.1\,:\,1} \\
    Echo + DTLN-AEC & 2.4\,\% & 24.0\,\% & 14.0\,\% & 59.6\,\% & 10.0\,:\,1 \\
    Dereverb     & 2.2\,\% & 4.2\,\%  & 4.4\,\%  & 89.2\,\% & 1.9\,:\,1  \\
    \hline
  \end{tabular}
  \caption{Decomposition of each condition's divergent clips against
    SLURP ground-truth intent (Whisper + Gemini~2.5 Flash Lite).
    \textbf{Correction}: clean prediction wrong, condition prediction
    correct. \textbf{Regression}: clean prediction correct, condition
    prediction wrong. \textbf{Both-Wrong}: both predictions wrong and
    mutually different. \textbf{Unaffected}: clean and condition
    predictions agree (correct or incorrect together). Regressions
    outnumber corrections in every condition, confirming ODR
    predominantly reflects harmful divergence rather than benign
    correction of clean-speech classification errors.}
  \label{tab:correction}
\end{table*}

\subsection{Quality Metrics as ODR Predictors}

Figure~\ref{fig:correlations} reports Pearson $r$ (left) and
Spearman $\rho$ (right) for each metric against both WER and ODR.

PESQ achieves the strongest Spearman correlation with ODR
($\rho\!=\!-0.467$), a moderate rank association at best.
The moderate ceiling is partly structural. Quality metrics quantify
signal-level fidelity while ODR measures semantic task fidelity,
dimensions that are only loosely coupled, particularly when failure
operates through categorical mechanisms such as speaker substitution
rather than continuous spectral degradation.
PESQ and STOI correlate comparably with WER and ODR
($\rho_{\text{WER}}\!=\!-0.439$ vs.\ $\rho_{\text{ODR}}\!=\!-0.467$
for PESQ), confirming that perceptual metrics track transcription
and semantic damage with similar (moderate) strength.
SQUIM-MOS, the most attractive non-intrusive estimator for production
monitoring (it requires no reference signal), is near-zero for both
($\rho_{\text{WER}}\!=\!-0.069$, $\rho_{\text{ODR}}\!=\!-0.068$),
ruling it out as a lightweight proxy for either target.
For echo specifically, AECMOS echo\_mos, a metric designed for
AEC quality assessment, achieves only $\rho\!=\!-0.080$ against
ODR, suggesting that even purpose-built echo quality metrics do not
reliably predict LLM task failure.

\paragraph{Pooled correlations overstate predictive value.}
The correlations above pool clips across conditions, conflating
within-condition variation with between-condition separation.
Echo (sim) has both the lowest mean PESQ and highest ODR as a
\emph{condition}, which alone can produce a substantial pooled
correlation even if PESQ does not distinguish diverged from
non-diverged clips \emph{within} a condition. Recomputing each
metric's correlation separately within each condition yields at most
$|\rho| = 0.34$ (STOI within MetricGAN{+}), with PESQ reaching
$|\rho| = 0.26$ and SNR, SRMR, and SQUIM-MOS below 0.15 (full
breakdown in Appendix~\ref{app:within_corr},
Table~\ref{tab:within_corr}), versus the pooled $\rho\!=\!-0.467$.
Thus, no metric provides meaningful signal about \emph{which} clips
will diverge once the enhancement condition is fixed.

This further shows that practitioners cannot use PESQ, or any of the
six tested metrics, to anticipate per-clip LLM failures within a given
SE pipeline. The pooled correlation reflects coarse condition-level
separation, not clip-level predictive power.

\paragraph{Where PESQ fails as an early-warning signal.}
Dichotomizing clips by within-condition median PESQ
(Table~\ref{tab:metricfail}) shows
that PESQ's failure mode is condition-dependent. Under Echo~(sim),
82.3\,\% of above-median-PESQ clips still diverge, meaning PESQ offers
no safety margin precisely when damage is worst. Under Dereverb and
Noisy, 84--88\,\% of below-median-PESQ clips do \emph{not} diverge,
meaning PESQ over-warns on clips the LLM handles correctly. No single
decision threshold on PESQ separates safe from unsafe clips across
conditions.

\begin{table}[t]
  \centering
  \resizebox{\columnwidth}{!}{%
  \begin{tabular}{lrrr}
    \hline
    \textbf{Condition} & \textbf{Med.\,PESQ} & \textbf{FN (\%)} & \textbf{FP(\%)} \\
    \hline
    Noisy        & 1.61 & 11.4 & \textbf{84.4} \\
    MetricGAN+   & 1.90 & 22.1 & 58.4 \\
    Echo (sim)   & 1.09 & \textbf{82.3} & 15.0 \\
    Echo + DTLN-AEC & 1.50 & 32.9 & 52.1 \\
    Dereverb     & 2.35 & 10.0 & \textbf{88.4} \\
    \hline
  \end{tabular}%
  }
  \caption{PESQ as a binary early-warning signal (threshold =
    within-condition median PESQ). FN: fraction of above-median clips
    that still diverge (PESQ misses real damage). FP: fraction of
    below-median clips that do not diverge (PESQ over-warns).
    No threshold separates safe from unsafe clips across conditions.}
  \label{tab:metricfail}
\end{table}

\subsection{Architecture Robustness}

Table~\ref{tab:model_comparison} (Appendix~\ref{app:arch}) confirms
that the findings are not Whisper-specific. wav2vec2 produces an
identical condition ranking (Spearman $\rho\!=\!1.0$, $p\!<\!0.001$),
with higher ODR under spectral distortion conditions
(mean $\Delta\!=\!+0.157$). The sole exception is Echo (sim), where
both models converge ($\Delta\!=\!+0.017$), confirming that
speaker-substitution failure is architecture-agnostic.

The higher wav2vec2 ODR under spectral conditions suggests that
CTC-based systems, common in latency-constrained deployments, may
face greater enhancement-induced LLM damage than encoder-decoder
estimates suggest.
 
\subsection{Domain-Level Vulnerability}
 
ODR varies across SLURP domains (Table~\ref{tab:domain_odr}).
Short formulaic commands
(\emph{lists}: mean ODR 0.419, \emph{alarm}: 0.386) are more
vulnerable than open-ended domains (\emph{news}: 0.319,
\emph{qa}: 0.315), as formulaic utterances have less lexical
redundancy for the LLM to recover intent from a degraded transcript.
Of 4{,}149 divergent clips, 90.4\% represent domain-level shifts
(predicted intent belongs to a different SLURP domain than the
ground-truth intent) and only 9.6\% are within-domain action
confusion, confirming that SE distortion destroys coarse domain
signals rather than producing fine-grained intent errors.
 
\subsection{LLM Robustness}
\label{sec:llm_robust}
 
To assess whether findings depend on classifier capacity, we
partially replicate the pipeline with Gemini~2.5 Pro.
On the full 2{,}974 clean and noisy clips, the direction of harm is
preserved: noisy ODR is 0.154 under Pro (vs.\ 0.135 under Flash Lite),
consistent with random variation at this scale.
Among the available 1{,}733 MetricGAN{+} clips, Pro ODR is 0.294.
The doubling direction is preserved: a substantially more capable LLM
still registers nearly twice as much semantic damage from
MetricGAN{+} as from unenhanced noise ($1.91{\times}$ vs.\
$2.36{\times}$ under Flash~Lite), consistent with the core finding
being robust to classifier capacity, though a full five-condition
Pro replication remains future work.
 
\section{Deployment Implications}
\label{sec:implications}

\paragraph{WER is an insufficient SE evaluation metric.}
Our results show WER can overestimate LLM damage (MetricGAN{+}:
ODR\,=\,0.318, Gap\,=\,$-0.296$) or saturate while
ODR remains catastrophic (Echo (sim): WER\,=\,1.448,
ODR\,=\,0.836).
We recommend ODR alongside WER as an offline evaluation criterion
for comparing SE configurations before deployment. Like PESQ and
STOI, it requires a clean reference recording and is therefore
a development-time tool rather than an online per-utterance monitor.
This recommendation carries urgency for CTC-based ASR
systems, where our results suggest ODR is systematically higher.
 
\paragraph{Echo cancellation is a hard requirement.}
The echo condition reaches an ODR of 0.836 and 0.853 under both ASR
models, meaning the spoken AI system correctly handles fewer than one in six
user interactions.
DNSMOS and PESQ score the near-end signal characteristics and cannot
detect this failure because they do not assess which speaker was
transcribed.
In any conferencing deployment, functional AEC must be treated as a
prerequisite for LLM integration, not an optional quality refinement
that can be deferred when latency or cost is constrained.

\paragraph{Perceptual metrics cannot replace LLM task evaluation.}
PESQ, MetricGAN{+}'s direct training objective, shows only moderate
rank association with ODR ($\rho = -0.467$), while non-intrusive
SQUIM-MOS shows near-zero association ($\rho = -0.068$).
SE pipelines optimised for perceptual quality provide no reliable
guarantee of downstream LLM performance. ODR measurement on a
representative intent task should be included in SE evaluation
protocols for spoken AI systems.
\section{Conclusion}
\label{sec:conclusion}

We introduced Output Divergence Rate (ODR), a task-level metric that
measures how often speech enhancement changes an LLM's intent
prediction relative to clean speech. Using ODR, we show that
perceptual quality improvements neither guarantee downstream LLM task
performance nor prevent degradation. MetricGAN{+} more than doubles
ODR despite improving PESQ, while unmitigated echo drives ODR to 0.836
through a distinct speaker-substitution failure beyond WER's scope.

Audio quality metrics range from near-zero to moderate correlation
with ODR and provide no reliable per-clip signal once the enhancement
condition is fixed. Both patterns replicate across two ASR architectures. Our
ground-truth intent decomposition further shows that SE-induced
divergence is predominantly harmful, rather than a benign side effect
of correcting clean-speech errors.

We recommend ODR alongside WER as an offline evaluation criterion
for SE front-ends deployed before an LLM.
The full pipeline is publicly released to support evaluation across
additional LLMs, tasks, and SE systems.
 
\section*{Limitations}

All five enhancement conditions in this study are simulated using DNS
Challenge corpora. Real device recordings may differ in acoustic
profile, compound distortion type, and room geometry, and we encourage
replication on field-recorded data.

On the ASR side, we compare encoder-decoder attention (Whisper) and
CTC (wav2vec2), but streaming CTC and attention-CTC hybrid architectures
remain untested. Given that wav2vec2 consistently yields higher ODR
than Whisper on spectral conditions, other CTC variants may amplify
the effects reported here.

The benchmark uses English-only SLURP data, and enhancement artefacts
may interact differently with ASR and LLM processing in tonal or
morphologically rich languages.

ODR is defined over a fixed 77-class intent vocabulary. Open-ended generation tasks (summarisation, question answering, or dialogue)
may exhibit different sensitivity to the transcript distortions induced
by each SE condition, and the mapping between our findings and those
settings requires further investigation.

Our primary results use Gemini~2.5 Flash Lite. Section~\ref{sec:llm_robust}
reports a partial Gemini~2.5 Pro replication.
Full multi-family LLM evaluation covering encoder-based classifiers
such as BERT-style joint intent/slot models, and extending the Pro
replication to all conditions, remains future work.

On the echo side, we evaluate a neural AEC system (DTLN-AEC).
Other architectures, including multi-microphone beamforming-based
cancellers and commercial AEC implementations, are untested and may
offer different recovery profiles.

Like PESQ and STOI, ODR requires a clean reference recording per clip,
which is typically unavailable during live deployment. It is therefore
intended as an offline, development-time criterion for comparing SE
configurations before deployment, not an online per-utterance monitor.
Reference-free proxies (transcript stability under paraphrase or ASR
confidence scores) are a natural direction for extending ODR to
deployment settings.

Among SE approaches, we test one discriminative GAN-based suppressor
(MetricGAN{+}), which is representative of deployed perceptual
optimisers. Diffusion-based, self-supervised, and proprietary commercial
SE systems remain untested, and the rate of intent collapse under those
approaches could differ substantially from what we report.

Finally, ODR is computed over closed-set intent labels only. The
SLURP test release used here does not expose entity or slot annotations,
so our error taxonomy does not separately track named-entity or
slot-level errors. Because SLURP intent labels are domain/action pairs
that often survive entity-level ASR substitutions (e.g.\ mishearing a
city name or contact), ODR may understate SE impact on tasks that
depend on entity fidelity such as slot filling or named-entity
recognition. Extending ODR to slot-level F\textsubscript{1} is future
work.

\section*{Ethical Considerations}
This work uses publicly available datasets (SLURP,
DNS Challenge) and publicly available models
(Whisper, MetricGAN{+}, WPE, wav2vec2). No
personally identifiable information was collected
or generated. The SLURP dataset was collected with
informed consent from participants. Our findings
expose a potential reliability failure in deployed
spoken AI systems, which we believe indicates that responsible
disclosure of such failures serves the public
interest by enabling practitioners to make
informed deployment decisions.
\vspace{-10pt}
\bibliography{custom}

 
\appendix
\section{Intent Classification Prompt}
\label{app:prompt}
 
The following prompt was held constant across all conditions, clips,
and ASR models.
 
\begin{quote}
\small\ttfamily
System: You are a spoken language understanding classifier.
Output ONLY one intent label from the list.
No punctuation. No explanation.
 
\medskip
User: Utterance: \{transcript\}
 
Valid intents: \{comma\_separated\_intent\_list\}
 
Intent:
\end{quote}
 
\section{ASR Architecture \& WER Comparison}
\label{app:arch}\label{app:wer}

Table~\ref{tab:model_comparison} provides the full per-condition
ODR and WER breakdown underlying the architecture comparison
in Section \ref{sec:results}.
\begin{table*}[t]
  \centering
  \begin{tabular}{llrrrrr}
    \hline
    \textbf{Condition} & \textbf{ASR} & \textbf{ODR} &
    \textbf{WER Mean} & \textbf{WER Med.} &
    \textbf{WER${>}$1.0} & \textbf{WER Max} \\
    \hline
    Echo (sim) & Whisper & \textbf{0.836} & 1.448 & 1.250 & 62.1\,\% & 9.0 \\
     & wav2vec2 & \textbf{0.853} & 1.472 & 1.300 & 66.1\,\% & 8.5 \\
    Echo + DTLN-AEC & Whisper & 0.404 & 0.700 & 0.625 & 9.8\,\% & 6.5 \\
     & wav2vec2 & 0.587 & 0.757 & 0.750 & 11.5\,\% & 5.0 \\
    MetricGAN+ & Whisper & 0.318 & 0.647 & 0.571 & 6.8\,\% & 5.0 \\
     & wav2vec2 & 0.474 & 0.624 & 0.600 & 5.1\,\% & 5.0 \\
    Noisy & Whisper & 0.135 & 0.531 & 0.500 & 2.4\,\% & 3.0 \\
     & wav2vec2 & 0.352 & 0.513 & 0.500 & 1.8\,\% & 4.5 \\
    Dereverb & Whisper & 0.108 & 0.517 & 0.500 & 1.7\,\% & 4.7 \\
     & wav2vec2 & 0.320 & 0.474 & 0.429 & 1.1\,\% & 4.0 \\
    \hline
    Clean & Whisper & -- & 0.507 & 0.444 & 1.2\,\% & 28.5 \\
     & wav2vec2 & -- & 0.388 & 0.333 & 0.3\,\% & 4.0 \\
    \hline
  \end{tabular}
  \caption{ODR and WER statistics across both ASR
    architectures (Gemini~2.5 Flash Lite for intent
    classification). \textbf{Bold} = highest ODR per model.
    Clean is shown as WER reference only (ODR\,=\,0 by
    definition). Values ${>}1.0$ arise when Whisper or
    wav2vec2 transcribes the far-end echo signal rather
    than the near-end speaker (Echo (sim)).}
  \label{tab:model_comparison}
\end{table*}

\section{Quality Metric Predictive Failure}
\label{app:metricfail}

See Table~\ref{tab:metricfail} in Section~\ref{sec:results}.

\vspace{-4pt}
\section{Domain-Level ODR}
\label{app:domain}

Table~\ref{tab:domain_odr} breaks down ODR by SLURP
semantic domain. Short-command domains (Lists, IoT, Alarm)
are most vulnerable, with mean ODR exceeding 0.38 across
conditions, while open-ended domains (QA, News, General)
are more robust. This suggests that enhancement-induced
lexical loss disproportionately affects utterances where a
single keyword carries the intent.
\begin{table*}[t]
  \centering
  \begin{tabular}{lrrrrrr}
    \hline
    \textbf{Domain} & \textbf{Noisy} & \textbf{MetricGAN+} & \textbf{Echo (sim)} & \textbf{Echo + DTLN-AEC} & \textbf{Dereverb} & \textbf{Mean} \\
    \hline
    Lists            & 0.201 & 0.429 & 0.886 & 0.429 & 0.150 & 0.419 \\
    IoT (WeMo)       & 0.074 & 0.444 & 0.889 & 0.526 & 0.148 & 0.416 \\
    Music            & 0.207 & 0.390 & 0.780 & 0.450 & 0.159 & 0.397 \\
    Social           & 0.162 & 0.371 & 0.886 & 0.417 & 0.114 & 0.390 \\
    Play             & 0.183 & 0.335 & 0.866 & 0.430 & 0.129 & 0.389 \\
    Calendar         & 0.154 & 0.343 & 0.879 & 0.427 & 0.128 & 0.386 \\
    Alarm            & 0.156 & 0.375 & 0.917 & 0.421 & 0.062 & 0.386 \\
    Volume           & 0.180 & 0.344 & 0.790 & 0.500 & 0.113 & 0.386 \\
    IoT (Hue)        & 0.167 & 0.339 & 0.874 & 0.390 & 0.143 & 0.382 \\
    Transport        & 0.153 & 0.323 & 0.847 & 0.419 & 0.113 & 0.371 \\
    \hline
    Email            & 0.124 & 0.311 & 0.891 & 0.417 & 0.090 & 0.366 \\
    Takeaway         & 0.019 & 0.333 & 0.944 & 0.300 & 0.130 & 0.345 \\
    Recommend.       & 0.075 & 0.237 & 0.892 & 0.413 & 0.075 & 0.338 \\
    Cooking          & 0.111 & 0.278 & 0.792 & 0.370 & 0.125 & 0.335 \\
    Datetime         & 0.070 & 0.280 & 0.830 & 0.400 & 0.070 & 0.330 \\
    IoT              & 0.131 & 0.328 & 0.770 & 0.321 & 0.098 & 0.330 \\
    Weather          & 0.086 & 0.237 & 0.815 & 0.404 & 0.079 & 0.324 \\
    News             & 0.057 & 0.262 & 0.754 & 0.481 & 0.041 & 0.319 \\
    QA               & 0.076 & 0.292 & 0.806 & 0.329 & 0.073 & 0.315 \\
    General          & 0.151 & 0.265 & 0.614 & 0.386 & 0.130 & 0.309 \\
    \hline
  \end{tabular}
  \caption{Per-domain mean ODR across all five enhancement conditions
    (Whisper + Gemini~2.5 Flash Lite), all 20 SLURP domains sorted by
    mean ODR descending. The horizontal rule separates domains with
    mean ODR $\geq 0.37$ (upper half) from those below. Short-command
    domains (Lists, IoT, Alarm) are consistently more vulnerable than
    open-ended domains (QA, News, General) across all conditions.}
  \label{tab:domain_odr}
\end{table*}

\section{Within-Condition Metric Correlations}
\label{app:within_corr}

\begin{table*}[!t]
  \centering
  \begin{tabular}{lrrrrr|r}
    \hline
    \textbf{Metric} & \textbf{Noisy} & \textbf{MetricGAN+} & \textbf{Echo (sim)} & \textbf{Echo+AEC} & \textbf{Dereverb} & \textbf{Pooled} \\
    \hline
    PESQ      & $-0.077$ & $\mathbf{-0.259}$ & $-0.048$ & $-0.202$ & $-0.036$ & $-0.467$ \\
    STOI      & $-0.107$ & $\mathbf{-0.340}$ & $-0.190$ & $-0.225$ & $-0.069$ & $-0.411$ \\
    SNR       & $+0.005$ & $-0.074$ & $+0.060$ & $\mathbf{+0.104}$ & $+0.007$ & $-0.116$ \\
    SI-SDR    & $-0.031$ & $\mathbf{-0.265}$ & $-0.012$ & $+0.073$ & $+0.007$ & $-0.066$ \\
    SRMR      & $-0.059$ & $\mathbf{-0.147}$ & $-0.023$ & $-0.061$ & $-0.063$ & $-0.111$ \\
    SQUIM-MOS & $-0.012$ & $\mathbf{-0.101}$ & $+0.036$ & $-0.085$ & $+0.025$ & $-0.068$ \\
    \hline
  \end{tabular}
  \caption{Within-condition Spearman $\rho$ (metric vs.\ binary diverged) for each
    enhancement condition, with the pooled cross-condition $\rho$ for comparison.
    \textbf{Bold} marks the largest $|\rho|$ per row.
    MetricGAN{+} is the only condition where any metric reaches $|\rho|>0.2$;
    STOI achieves the maximum ($-0.340$), PESQ reaches $-0.259$.
    Positive SNR and SI-SDR values within Echo + DTLN-AEC reflect a sign
    inversion: the nonlinear gain and phase modifications applied by the AEC
    break the monotone fidelity-divergence relationship, consistent with the
    SI-SDR anomaly in Table~\ref{tab:metrics}.
    The large pooled-to-within drop confirms that pooled correlations reflect
    between-condition separation, not clip-level predictive power.}
  \label{tab:within_corr}
\end{table*}

Table~\ref{tab:within_corr} reports Spearman $\rho$ between each
quality metric and clip-level divergence computed \emph{separately}
within each enhancement condition, alongside the pooled cross-condition
value for comparison.
Within any single condition, no metric exceeds $|\rho| = 0.34$
(STOI within MetricGAN{+}). Most are below 0.15, and several change
sign across conditions.
The large reduction from pooled to within-condition values confirms
that the pooled correlations (Table~\ref{tab:metrics},
Figure~\ref{fig:correlations}) reflect between-condition separation
rather than clip-level predictive power.

\section{Log-Mel Spectrogram Comparison Across Conditions}
\label{app:spectrograms}

Figure~\ref{fig:spectrograms} shows log-mel spectrograms of five
representative SLURP utterances across all six conditions.
Each panel shares the same colour scale (amplitude) and axis
conventions: time on the horizontal axis, frequency (kHz) on the
vertical axis.
Panels in the same row depict the same utterance. The ground-truth
transcript appears as a centred heading above each row.

\definecolor{cclean}{HTML}{000000}
\definecolor{cnoisy}{HTML}{000000}
\definecolor{cmg}{HTML}{000000}
\definecolor{caec}{HTML}{000000}
\definecolor{caecf}{HTML}{000000}
\definecolor{cdereverb}{HTML}{000000}

\newlength{\specw}
\setlength{\specw}{2.2cm}

\begin{figure*}[p]
\centering
\setlength{\tabcolsep}{1pt}
\renewcommand{\arraystretch}{0.5}
\begin{tabular}{cccccc}
\noalign{\vspace{3pt}}\hline\noalign{\vspace{3pt}}
  \textbf{\textcolor{cclean}{Clean}} &
  \textbf{\textcolor{cnoisy}{Noisy}} &
  \textbf{\textcolor{cmg}{MetricGAN{+}}} &
  \textbf{\textcolor{caec}{Echo (sim)}} &
  \textbf{\textcolor{caecf}{Echo + DTLN-AEC}} &
  \textbf{\textcolor{cdereverb}{Dereverb}} \\[2pt]
\noalign{\vspace{3pt}}\hline\noalign{\vspace{3pt}}
  \multicolumn{6}{c}{\textit{``I want to hear snow by red hot chili peppers''}}\\
  \includegraphics[height=1.5\specw]{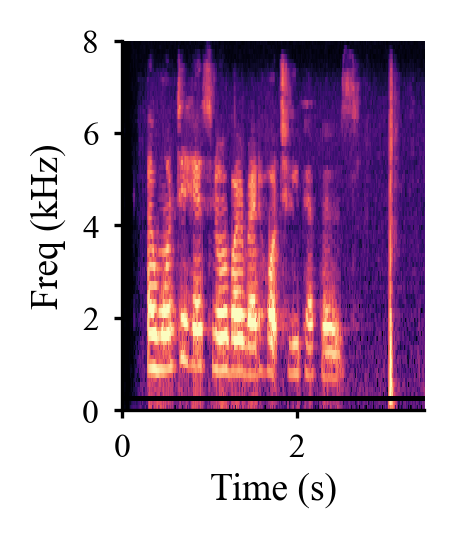} &
  \includegraphics[height=1.5\specw]{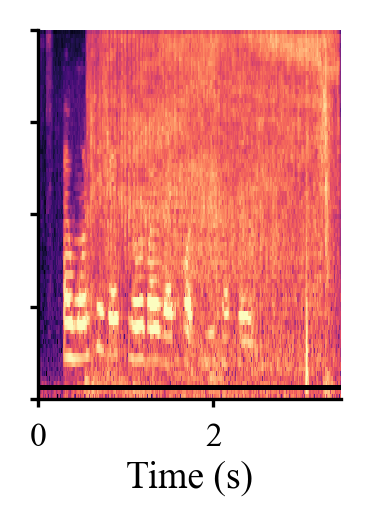} &
  \includegraphics[height=1.5\specw]{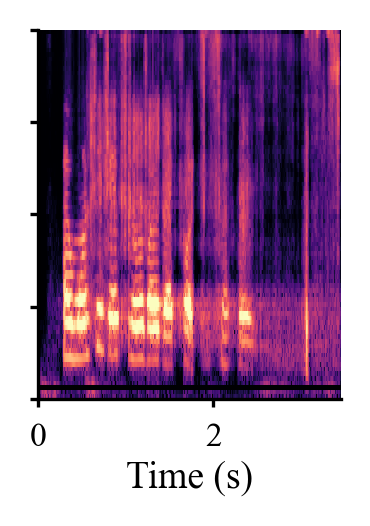} &
  \includegraphics[height=1.5\specw]{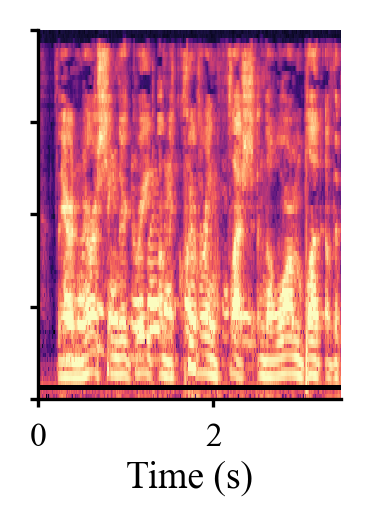} &
  \includegraphics[height=1.5\specw]{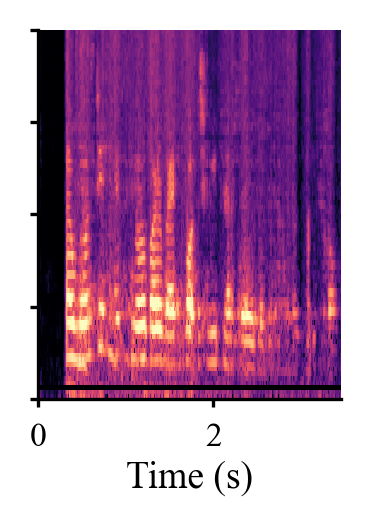} &
  \includegraphics[height=1.5\specw]{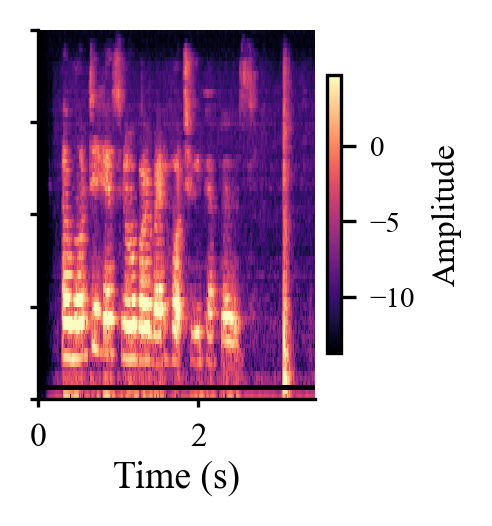} \\
\noalign{\vspace{3pt}}\hline\noalign{\vspace{3pt}}
  \multicolumn{6}{c}{\textit{``How old are you''}}\\
  \includegraphics[height=1.5\specw]{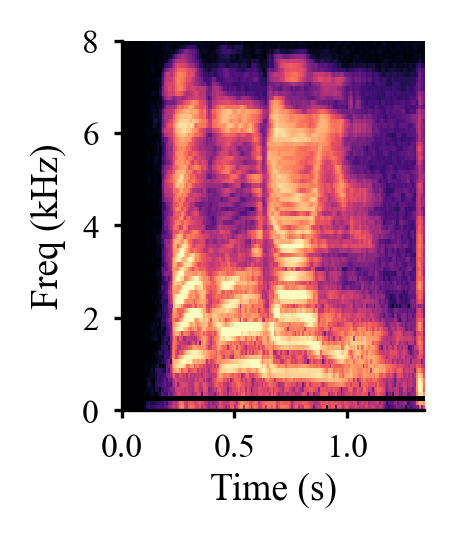} &
  \includegraphics[height=1.5\specw]{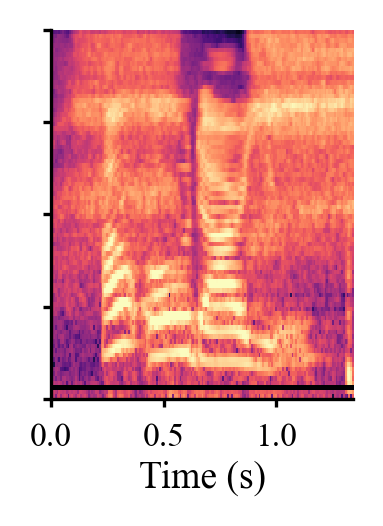} &
  \includegraphics[height=1.5\specw]{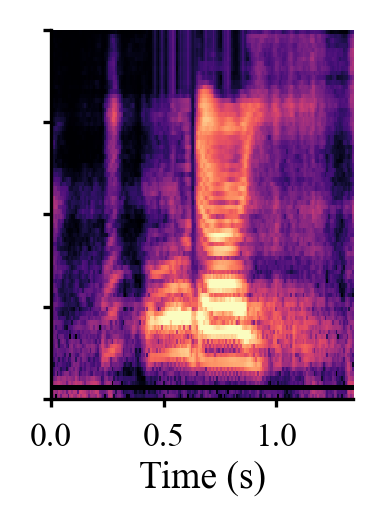} &
  \includegraphics[height=1.5\specw]{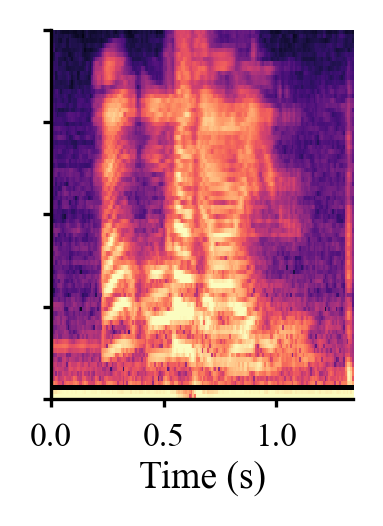} &
  \includegraphics[height=1.5\specw]{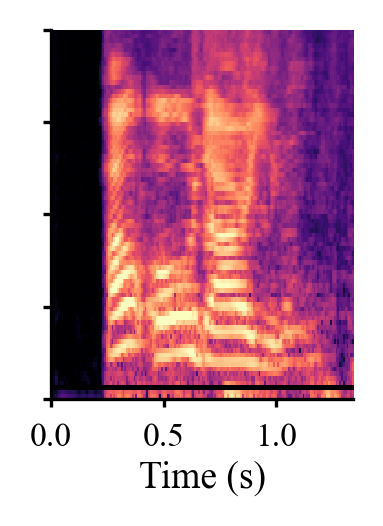} &
  \includegraphics[height=1.5\specw]{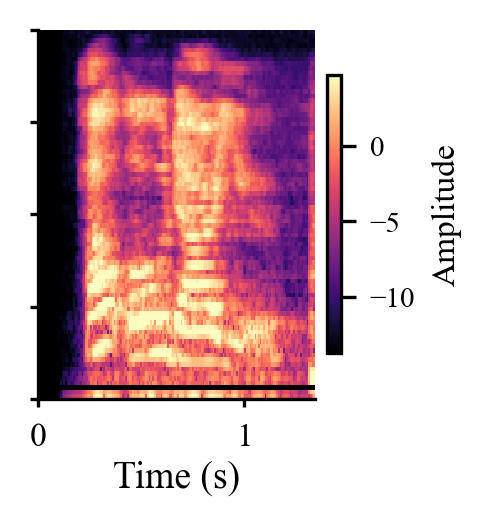} \\
\noalign{\vspace{3pt}}\hline\noalign{\vspace{3pt}}
  \multicolumn{6}{c}{\textit{``Weather report now''}}\\
  \includegraphics[height=1.5\specw]{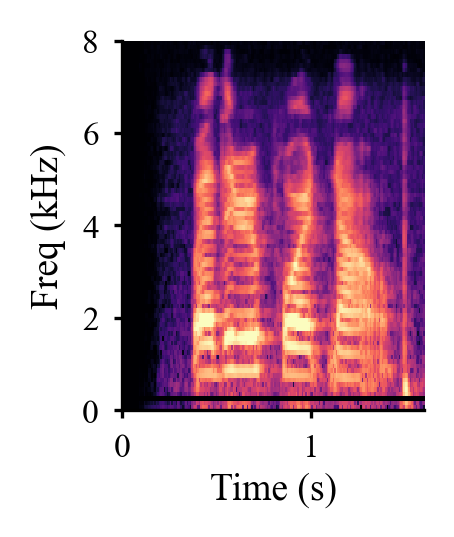} &
  \includegraphics[height=1.5\specw]{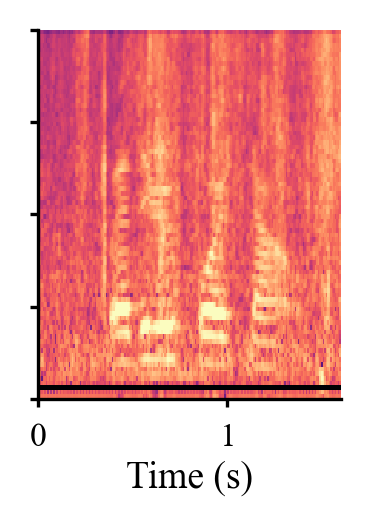} &
  \includegraphics[height=1.5\specw]{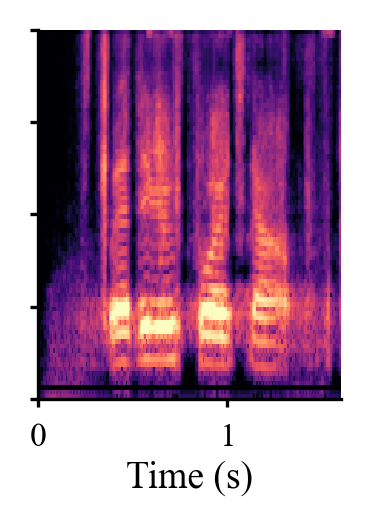} &
  \includegraphics[height=1.5\specw]{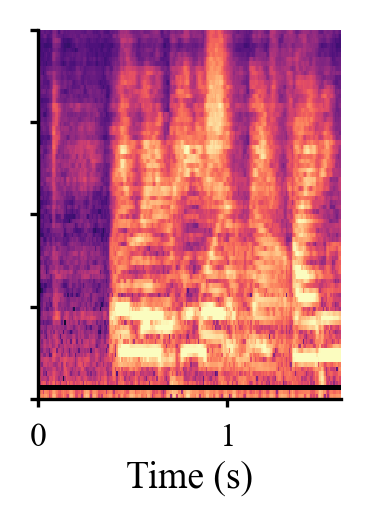} &
  \includegraphics[height=1.5\specw]{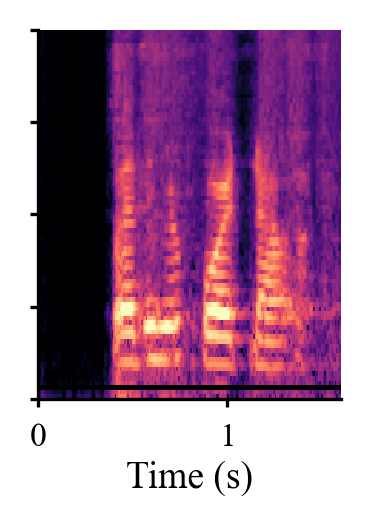} &
  \includegraphics[height=1.5\specw]{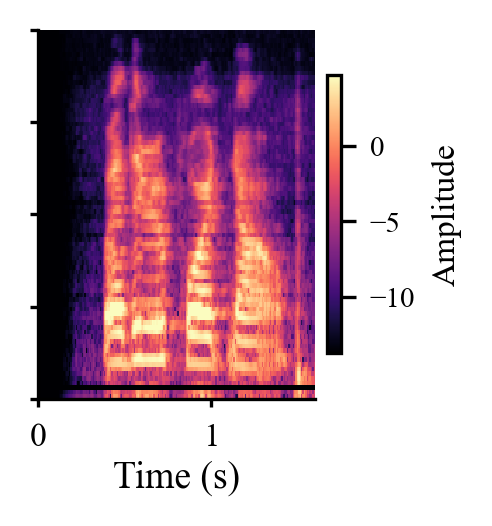} \\
\noalign{\vspace{3pt}}\hline\noalign{\vspace{3pt}}
  \multicolumn{6}{c}{\textit{``I would like it to help analyze ideas''}}\\
  \includegraphics[height=1.5\specw]{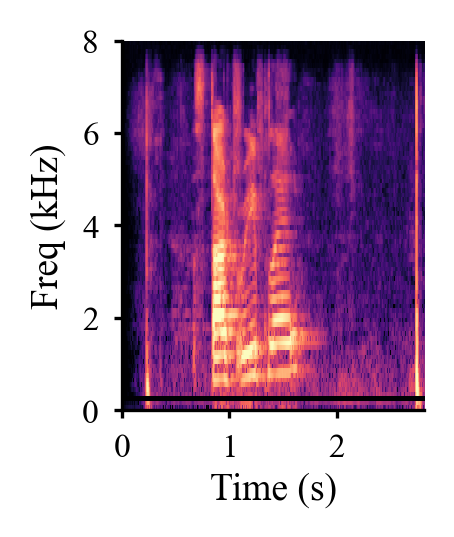} &
  \includegraphics[height=1.5\specw]{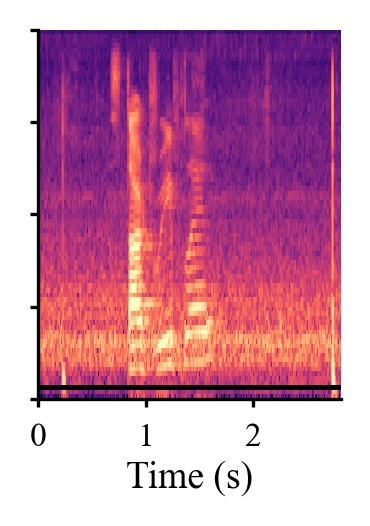} &
  \includegraphics[height=1.5\specw]{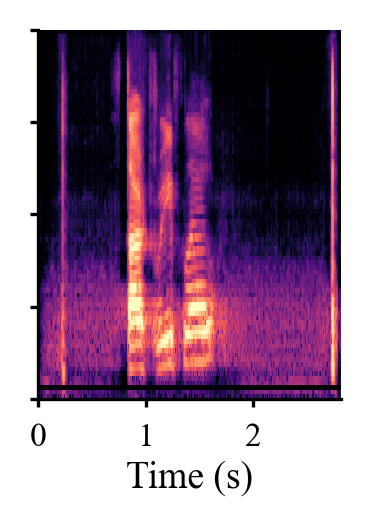} &
  \includegraphics[height=1.5\specw]{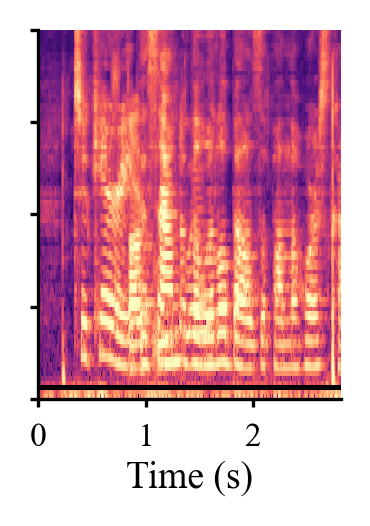} &
  \includegraphics[height=1.5\specw]{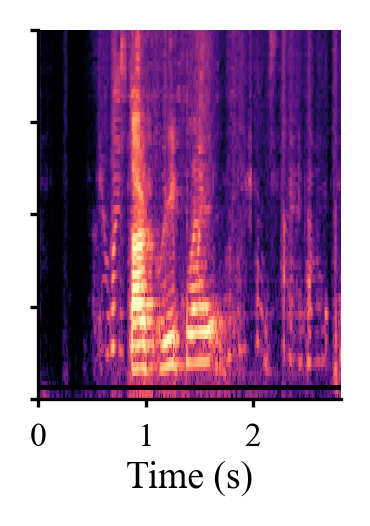} &
  \includegraphics[height=1.5\specw]{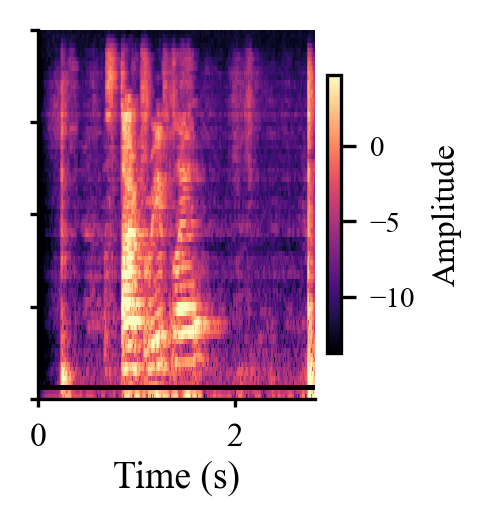} \\
\noalign{\vspace{3pt}}\hline\noalign{\vspace{3pt}}
  \multicolumn{6}{c}{\textit{``I'm going to sleep now''}}\\
  \includegraphics[height=1.5\specw]{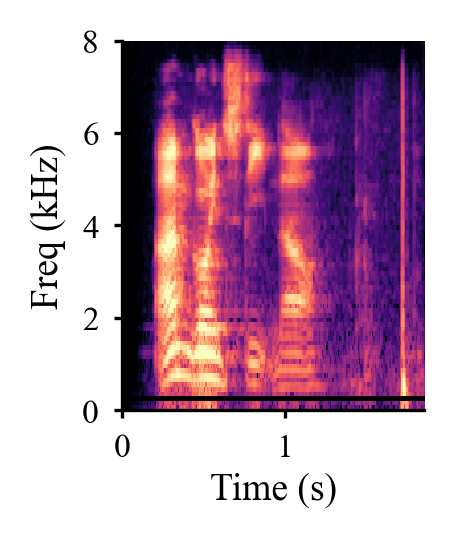} &
  \includegraphics[height=1.5\specw]{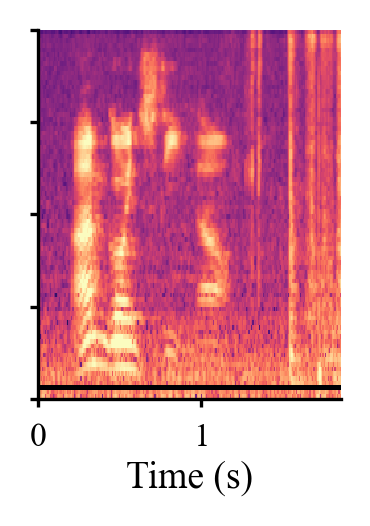} &
  \includegraphics[height=1.5\specw]{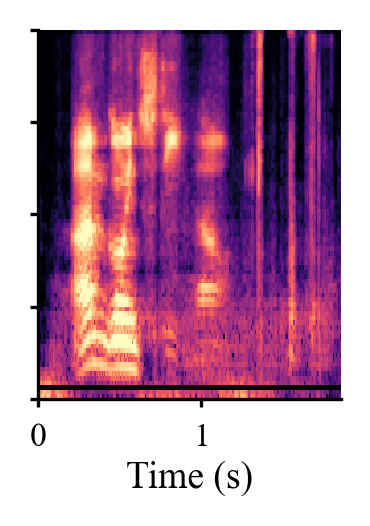} &
  \includegraphics[height=1.5\specw]{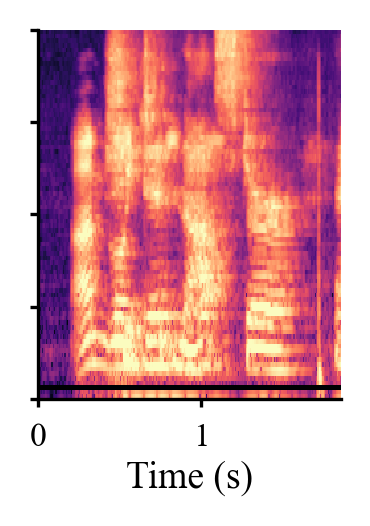} &
  \includegraphics[height=1.5\specw]{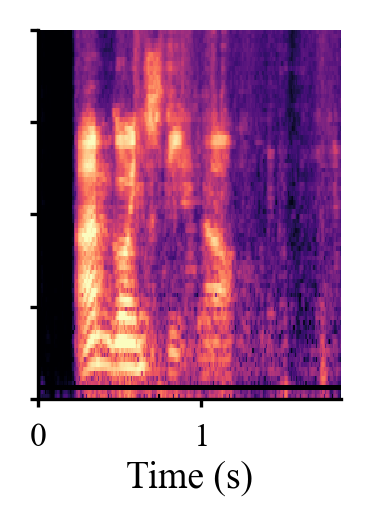} &
  \includegraphics[height=1.5\specw]{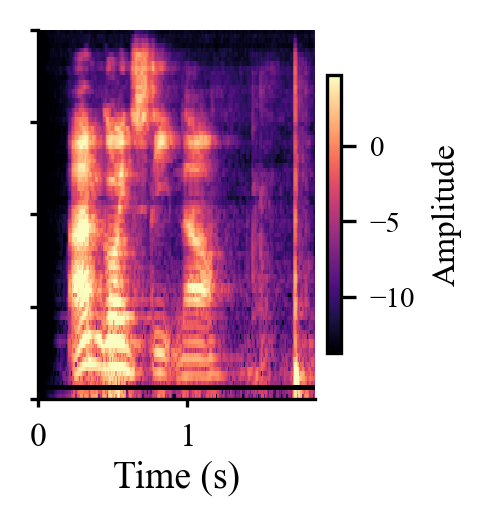} \\
\noalign{\vspace{3pt}}\hline
\end{tabular}
\caption{Log-mel spectrograms of five SLURP utterances (rows) across
  all six conditions (columns). Horizontal axis: time~(s);
  vertical axis: frequency~(kHz); colour: amplitude (log-mel energy,
  shared scale across all panels).
  The ground-truth transcript of each utterance appears as a centred
  heading above its row.
  Noisy floods the high-frequency band with broadband noise.
  MetricGAN{+} suppresses it but introduces vertical streak artefacts
  (musical noise, visible above 4\,kHz) and alters the harmonic envelope.
  Echo~(sim) shows a wholly different far-end utterance dominating the
  near-end speech.
  Echo~+~DTLN-AEC partially restores the near-end harmonic structure.
  Dereverb is visually closest to Clean, consistent with its low ODR.
  Row~5 (``I'm going to sleep now'') is a robust clip that does not
  diverge under any condition; all six panels remain visually similar.}
\label{fig:spectrograms}
\end{figure*}

\vspace*{\fill}

\end{document}